\documentclass[amsmath,amssymb,aps,pre,reprint, showpacs, superscriptaddress]{revtex4-1}
\usepackage{graphicx}
\usepackage{amsthm}
\usepackage{amsmath}
\usepackage{amssymb}
\usepackage{dsfont}
\usepackage{bm}

\usepackage{color}
\begin{document} 
\title[]{Efficiency at and near Maximum Power of Low-Dissipation Heat Engines}
\author{Viktor Holubec}
\email{viktor.holubec@mff.cuni.cz}
\affiliation{ 
 Charles University in Prague,  
 Faculty of Mathematics and Physics, 
 Department of Macromolecular Physics, 
 V Hole{\v s}ovi{\v c}k{\' a}ch 2, 
 CZ-180~00~Praha, Czech Republic 
}
\author{Artem Ryabov}
\affiliation{ 
 Charles University in Prague,  
 Faculty of Mathematics and Physics, 
 Department of Macromolecular Physics, 
 V Hole{\v s}ovi{\v c}k{\' a}ch 2, 
 CZ-180~00~Praha, Czech Republic 
}
\date{\today} 
%%%%%%%%%%%%%%%%%%%%%%%%%%%%%%%%%%%%%%%%%%%%%%%%%%%%%%%%%%%%%%%%%%%%%%%%%%%%%%%%%%%%%%%%%%%%%%%%%%%%%%%%%%%%%%%%%%
\begin{abstract} 
A new universality in optimization of trade-off between power and efficiency for low-dissipation Carnot cycles is presented. It is shown that any trade-off measure expressible in terms of efficiency and the ratio of power to its maximum value can be optimized independently of most details of the dynamics and of the coupling to thermal reservoirs. The result is demonstrated on two specific trade-off measures. The first one is designed for finding optimal efficiency for a given output power and clearly reveals diseconomy of engines working at maximum power. 
As the second example we derive universal lower and upper bounds on the efficiency at maximum trade-off given by the product of power and efficiency. The results are illustrated on a model of a diffusion-based heat engine. Such engines operate in the low-dissipation regime given that the used driving minimizes the work dissipated during the isothermal branches. The peculiarities of the corresponding optimization procedure are reviewed and thoroughly discussed.
\end{abstract}

\pacs{05.20.-y, 05.70.Ln, 07.20.Pe} 
% Mechanics statistical, 05.20.-y
% Irreversible thermodynamics, 05.70.Ln
% Heat engines, 07.20.Pe

\maketitle  
%%%%%%%%%%%%%%%%%%%%%%%%%%%%%%%%%%%%%%%%%%%%%%%%%%%%%%%%%%%%%%%%%%%%%%%%%%%%%%%%%%%%%%%%%%%%%%%%%%%%%%%%%%%%%%%%%%
%%%%%%%%%%%%%%%%%%%%%%%%%%%%%%%%%%%%%%%%%%%%%%%%%%%%%%%%%%%%%%%%%%%%%%%%%%%%%%%%%%%%%%%%%%%%%%%%%%%%%%%%%%%%%%%%%%
%%%%%%%%%%%%%%%%%%%%%%%%%%%%%%%%%%%%%%%%%%%%%%%%%%%%%%%%%%%%%%%%%%%%%%%%%%%%%%%%%%%%%%%%%%%%%%%%%%%%%%%%%%%%%%%%%%
\section{Introduction}

Since the discovery of heat engines people were struggling to optimize their performance \cite{Muller2007}. One of the first results in the field was due to Sadi Carnot \cite{Carnot1978} and Rudolf Clausius \cite{Clausius1856}. They showed that the maximum efficiency attainable by any heat engine operating between the hot reservoir at the temperature $T_{\rm h}$ and the cold one at the temperature $T_{\rm c}$ is given by the Carnot efficiency $\eta_{\rm C} = 1 - T_{\rm c}/T_{\rm h}$. Alas their result was of limited practical importance for the corresponding heat engine must work reversibly and thus its output power is zero. Optimization of the power of \emph{irreversible} Carnot cycles working under finite-time conditions turned out to be much more practical. It was pioneered by Novikov \cite{Novikov1958}, Chambadal \cite{Chambadal1957} and later by Curzon and Ahlborn \cite{Curzon1975}. The obtained result $\eta_{\rm CA} = 1 - \sqrt{T_{\rm c}/T_{\rm h}}$ for the efficiency at maximum power (EMP) seemingly exhibits the same degree of generality as the Carnot's formula and also it describes well the efficiency of some actual thermal plants \cite{Callen2006, Curzon1975, Bejan1997}. Although it turned out that $\eta_{\rm CA}$ is not a universal result, neither it represents an upper or lower bound for the EMP \cite{Hoffmann1997, Berry2000, Salamon2001}, its close agreement with EMP for several model systems \cite{DeVos1985,Bejan1996,JimenezdeCisneros2007,Schmiedl2008,Izumida2008,Izumida2009,Allahverdyan2008,Tu2008,Esposito2009a,Rutten2009,Esposito2010, Zhou2010,Zhan-Chun2012} ignited search for universalities in performance of heat engines.

For thermochemical engines it was shown that the form of the EMP \cite{Esposito2009} is controlled by the symmetries of the underlying dynamics (see also \cite{Izumida2014} for a similar study, but with finite-sized reservoirs). This result was further precised \cite{Sheng2015} and recently a similar argumentation was successfully used for general thermodynamic devices \cite{Cleuren2015}. Further universalities were obtained for the class of heat engines working in the regime of ``low dissipation'' \cite{Esposito2010b, Sekimoto1997, Bonanca2014, Schmiedl2008, Tomas2013, Muratore-Ginanneschi2015}, where the work dissipated during the isothermal branches of the Carnot cycle grows in inverse proportion to the duration of these branches. Namely a general expression for the EMP for these engines has been published \cite{Schmiedl2008} and subsequently lower and upper bounds on the EMP  were derived \cite{Esposito2010b}. All these results have been confirmed within a minimally nonlinear irreversible thermodynamics framework \cite{Izumida2012, Izumida2013}. Another universal result recently appeared in the realm of Brownian motors \cite{Astumian2002, Hanggi2009, Hanggi2005, Seifert2012}. Engines so small that their output work and input heat become stochastic due to thermal fluctuations. The ratio of these variables yields the efficiency which is thus itself random variable. Large deviation form of the corresponding probability distribution exhibits the following universal features \cite{Verley2014,Verley2014a, Rana2014,Rana2015, Proesmans2015, Polettini2015, Martinez2014, Gingrich2014, Esposito2015}: (i) the most probable value of efficiency corresponds to its traditional definition using averages of work and heat, (ii)
in case of time-reversal symmetric driving the least probable value of efficiency equals to $\eta_{\rm C}$.

In the present paper we reveal a new universality in optimization of \emph{trade-off between power and efficiency} for low-dissipation Carnot cycles \cite{Tomas2013, Long2015, Long2014, Sheng2013}. As our main result, we show that any trade-off measure expressible in terms of efficiency and the ratio of power to its maximum value exhibits the same degree of universality as the EMP (see Sec.~\ref{sec:tradeoff}). Consequently, for many systems these trade-off measures can be optimized independently of most details of the dynamics and of the coupling to thermal reservoirs. Investigation of the trade-off between power and efficiency is immensely important for engineering practice, where not only powerful, but also economical devices should be developed. Indeed, it was already highlighted \cite{Chen2001,DeVos1992,Chen1994} that actual thermal plants and heat engines should not work at the maximum power, where the corresponding efficiency can be relatively small, but rather in a regime with slightly smaller power and considerably larger efficiency. Our results strongly support these findings.

Examples of the low-dissipation Carnot cycles are Brownian motors based on two-level quantum systems \cite{Zulkowski2015a} or on driven overdamped systems \cite{Schmiedl2008,Zulkowski2015}, given that the used driving minimizes the work dissipated during the isothermal branches and that all thermodynamic variables are defined in the traditional manner, i.e., via averages. We illustrate our results using Carnot cycles based on an overdamped particle diffusing in an externally controlled confining potential. Due to the recent advances in micromanipulation techniques \cite{Ritort2008}, such engines were already realized experimentally \cite{Martinez2014, 
Blickle2011} and hence are of vital theoretical interest \cite{Schmiedl2008, Holubec2014, Rana2014, Rana2015,Benjamin2008,Tu2014}. 

The paper is organized as follows. In Sec.~\ref{sec:model} we introduce the diffusion-based heat engines and standard thermodynamic variables used to characterize their performance \cite{Schmiedl2008, Seifert2012}. The scheme \cite{Schmiedl2007,Schmiedl2008} for minimization of the work dissipated during the isothermal branches which leads to the low-dissipation behavior is described in Sec.~\ref{sec:max_eta}. In Sec.~\ref{sec:maxP}, we review the results \cite{Schmiedl2008, Esposito2010b} concerning the EMP for low-dissipation Carnot cycles. In the both sections we discuss consequences of the optimisation procedure on specific model systems. Our main conclusions are: (i) Under the restrictions of the optimization procedure, the protocol which minimizes the dissipation yields, at the same time, the maximum power and efficiency. (ii) When the exact optimal protocol can not be obtained, a numerical optimization within a certain class of ``test'' protocols yields performance comparable to that of the optimal protocol. (iii) The scheme for minimization of the work dissipated during the isothermal branches leads to the low-dissipation behavior only in case the optimization procedure has enough freedom to modify the potential.

In Sec.~\ref{sec:tradeoff} we utilize the findings on the EMP reviewed in Sec.~\ref{sec:maxP} to demonstrate the universality in the trade-off between power and efficiency. As particular examples we optimize two specific trade-off measures. The first one (\ref{eq:tradeoff}) is suitable for finding optimal efficiency for a given output power. The results of the optimization procedure are illustrated on the model of log-harmonic potential 
introduced in Secs.~\ref{sec:model}-\ref{sec:max_eta}. The performed optimization indeed shows that engines operating near maximum power exhibit only slightly smaller power than the maximal one, but, at the same time, work with considerably larger efficiency than the EMP. Within the second example we derive universal bounds (\ref{eq:BoundsMV}) on the efficiency at maximum trade-off (\ref{eq:Lambda}) given by the product of power and efficiency. By comparison with data from actual nuclear power plants these bounds give promising results.

%%%%%%%%%%%%%%%%%%%%%%%%%%%%%%%%%%%%%%%%%%%%%%%%%%%%%%%%%%%%%%%%%%%%%%%%%%%%%%%%%%%%%%%%%%%%%%%%%%%%%%%%%%%%%%%%%%
\section{Diffusion-based heat engines}
%%%%%%%%%%%%%%%%%%%%%%%%%%%%%%%%%%%%%%%%%%%%%%%%%%%%%%%%%%%%%%%%%%%%%%%%%%%%%%%%%%%%%%%%%%%%%%%%%%%%%%%%%%%%%%%%%
\label{sec:model}

\begin{figure}
	\centering
		\includegraphics[width=1.0\columnwidth]{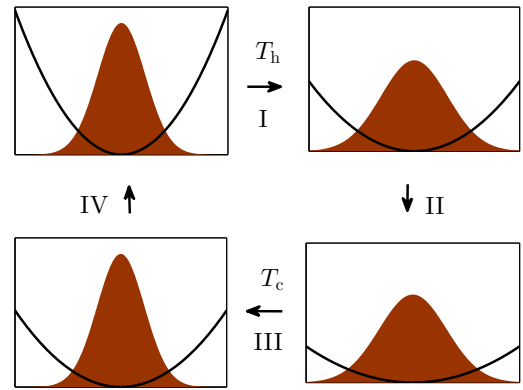}
	\caption{(Color
  online) Sketch of a diffusion-based heat engine. The periodic changes of the potential (black line) and of the reservoir temperature $T(t)$ induce the periodic response of the PDF to find the particle at time $t$ at position $x$ (filled brown curve.)}
	\label{fig:cycle}
\end{figure}

Consider the Carnot-like model \cite{Schmiedl2008, Holubec2014} sketched in Fig.~\ref{fig:cycle}. The working medium of the heat engine is an overdamped particle diffusing in an externally driven confining potential. Both the potential, $U(x,t)$, and the bath temperature, $T(t)$, are periodically altered. The cycle consists of two isotherms (branches I and III) and two adiabats (branches II and IV).

State of the engine is described by the probability density function (PDF) to find the particle at time $t$ at position $x$. Under the time-periodic driving the engine will eventually attain a time-periodic steady state, the so called limit cycle. The corresponding PDF $p(x,t)$ follows from the Fokker-Planck equation
\begin{align}
\frac{\partial}{\partial t}p(x,t) &= - \frac{\partial}{\partial x} j(x,t)\,,
\label{eq:Fokker-Planck}\\
j(x,t) &= - \mu(t)\left\{ k_{\rm B}T(t)\frac{\partial}{\partial x} +  \left[\frac{\partial U(x,t)}{\partial x}\right]  \right\}p(x,t)\,,
\label{eq:current}
\end{align}
supplemented by the condition 
\begin{equation}
p(x,t + t_{\rm p}) = p(x,t)\,.
\label{eq:limit_cycle}
\end{equation}
Here $\mu(t)$ stands for the mobility, which is assumed to depend on temperature, $k_{\rm B}$ is the Boltzmann constant, $j(x,t)$ denotes the probability current and $t_{\rm p}$ is the duration of one operational cycle of the engine. 

In the following subsections \ref{subs:IIA} and \ref{subs:IIB} we will specify the individual branches of the heat engine. In subsection \ref{subs:IIC} we will discuss on examples under which conditions the heat engine is capable of producing positive work. Other performance characteristics of the engine like its efficiency and output power will be introduced in section~\ref{sec:max_eta} and discussed in the rest of the paper.

%%%%%%%%%%%%%%%%%%%%%%%%%%%%%%%%%%%%%%%%%%%%%%%%%
\begin{table*}%The best place to locate the table environment is directly after its first reference in text
\caption{\label{tab:potential_state}%
Potentials for which periodic solutions $p(x,t)$ of the Fokker-Planck equation~(\ref{eq:Fokker-Planck}) are known. The engine based on the
sliding parabola (the second line) was studied in
\cite{Holubec2014a, Mazonka1999}, the engine based on the breathing parabola (the third line) is discussed in \cite{Schmiedl2008} and the engine based on the log-harmonic potential (the fourth line) was introduced in \cite{Holubec2014}. The response functions $m(t)$, $\sigma(t)$ and $f(t)$ are continuous and depend on the specific form of the driving. The irreversible works $W^{\rm irr}_{\rm I, \rm III}$ (\ref{eq:irreversible_work}) for the sliding parabola are functionals of the mean particle position $m(t)$. The variance $\sigma(t)$ enters only the boundary terms $g_{\rm I, \rm III} = \mp k\left[\sigma(t_{\rm h}) - \sigma(0)\right]/2 \pm T_{\rm I, \rm III}\Delta S$. The irreversible works for the breathing parabola and for the log-harmonic potential are functionals of the variances $\sigma(t)$ and $4(\nu+1)f(t)$ respectively. In the last column we show optimized irreversible works for the optimal responses (\ref{eq:mopt})-(\ref{eq:fopt}).
}
\begin{ruledtabular}
\begin{tabular}{l|l|l|l}
$U(x,t)$& $p(x,t)$& $W^{\rm irr}_{\rm I, \rm III}$& optimized $W^{\rm irr}_{\rm I, \rm III} = A_{\rm I, \rm III}/t_{\rm I, \rm III}$\\[1pt]
\colrule
\colrule
$\frac{1}{2}k[x - y(t)]^2$ &  $\frac{1}{\sqrt{2\pi \sigma(t)}}\exp\left\{-\frac{[x-m(t)]^2}{2\sigma(t)}\right\}$ & $g_{\rm I, \rm III} + \int_{\rm I, \rm III} {d}t\,\frac{1}{\mu(t)}\left[\dot{m}(t)\right]^2$ &  $g_{\rm I, \rm III} + \frac{1}{\mu_{\rm I, \rm III}t_{\rm I, \rm III}}[m(t_{\rm h}) - m(0)]^2$\\[1pt]
$\frac{1}{2}k(t)x^2$ &  $\frac{1}{\sqrt{2\pi \sigma(t)}}\exp\left[-\frac{x^2}{2\sigma(t)}\right]$ & $\frac{1}{4}\int_{\rm I, \rm III} {d}t\,\frac{1}{\mu(t)\sigma(t)}\left[\dot{\sigma}(t)\right]^2$ & $\frac{1}{\mu_{\rm I, \rm III}t_{\rm I, \rm III}}(\sqrt{\sigma(t_{\rm h})} - \sqrt{\sigma(0)})^2$\\[1pt]
$-(2\nu+1)T(t)\ln |x|+\frac{1}{2}k(t)x^2$  &  $\frac{[f(t)]^{-\nu-1}}{\Gamma(\nu+1)}\left(\frac{x}{2}\right)^{2\nu+1}\exp\left[-\frac{x^2}{4f(t)}\right]$ & $\int_{\rm I, \rm III} {d}t\,\frac{(\nu+1)}{\mu(t)f(t)}\left[\dot{f}(t)\right]^2$ & $\frac{4(\nu+1)}{\mu_{\rm I, \rm III} t_{\rm I, \rm III}}(\sqrt{f(t_{\rm h})} - \sqrt{f(0)})^2$\\
\end{tabular}
\end{ruledtabular}
\end{table*}
%%%%%%%%%%%%%%%%%%%%%%%%%%%%%%%%%%%%%%%%%%%%%%%%%

%%%%%%%%%%%%%%%%%%%%%%%%%%%%%%%%%%%%%%%%%%%%%%%%%%%%%%%%%%%%%%%%%%%%%%%%%%%%%%%%%%%%%%%%%%%%%%%%%%
\subsection{Branches I and III: isotherms}
\label{subs:IIA}
%%%%%%%%%%%%%%%%%%%%%%%%%%%%%%%%%%%%%%%%%%%%%%%%%%%%%%%%%%%%%%%%%%%%%%%%%%%%%%%%%%%%%%%%%%%%%%%%%%

During the isothermal branches the system is in contact with heat reservoir at constant temperature which we denote as $T_{\rm I} = T_{\rm h}$ for the hot isotherm (branch I in Fig.~\ref{fig:cycle}) and $T_{\rm III} = T_{\rm c}$ for the cold one (branch III in Fig.~\ref{fig:cycle}). Hence the mobility will be a piece-wise constant function. We denote its values during the hot and cold isotherms as $\mu_{\rm I}=\mu_{\rm h}$ and $\mu_{\rm III}=\mu_{\rm c}$ respectively. Along the isotherms the potential and also the PDF changes with time. We denote as $t_{\rm I} = t_{\rm h}$ and $t_{\rm III} = t_{\rm c}$ the durations of the hot and of the cold isotherms respectively.

The work produced by the engine during the isotherms can be written as \cite{Seifert2012, Schmiedl2008}
\begin{multline}
W_{\rm I, \rm III} = -\int_{{\rm I},{\rm III}}{d}t\int{d}x\,p(x,t)\frac{\partial }{\partial t} U(x,t)=\\
=  - W^{\rm irr}_{\rm I, \rm III} - \Delta E_{\rm I, \rm III} + T_{\rm I, \rm III} \Delta S_{\rm I, \rm III} \,,
\label{eq:isothermal_work}
\end{multline}
where $\int_{{\rm I}}$ denotes that the integration should be carried out along the first branch (and similarly for $\int_{{\rm III}}$). The symbols $\Delta E_{\rm I, \rm III}$ stand for the \emph{increase} of (mean) internal energy of the system during the individual branches and 
\begin{multline}
\Delta S = \Delta S_{\rm I} = - \Delta S_{\rm III} =\\
= - k_{\rm B} \int{d}x \left[ p(x,t_{\rm h}) \ln p(x,t_{\rm h}) - p(x,0) \ln p(x,0) \right] 
\label{eq:system_entropy}
\end{multline}
is the increase of system entropy during the branch I. 

The last two terms on the right-hand side of the formula (\ref{eq:isothermal_work}) combine into the decrease of (non-equilibrium) free energy. Therefore, according to the second law of thermodynamics, for an equilibrium process $W^{\rm irr}_{\rm I, \rm III}$ vanish. These quantities measure the amount of work which was dissipated during the individual isothermal branches and must always be positive. They are known as \emph{irreversible works} \cite{Schmiedl2008} and can be calculated as \cite{Tome2006}
\begin{equation}
W^{\rm irr}_{\rm I, \rm III} = \frac{1}{t_{\rm I, \rm III}}\frac{1}{\mu_{\rm I, \rm III}} \int_{0}^{1}{d}\tau\int{d}x\,
\frac{[J_{\rm I, \rm III}(x,\tau)]^2}{P_{\rm I, \rm III}(x,\tau)}\ge 0\,.
\label{eq:irreversible_work}
\end{equation}
The quantities in the denominator of the integrand are time-rescaled probability densities $P_{\rm I}(x,\tau) = p(x,t_{\rm h}\tau)$ and $P_{\rm III}(x,\tau) = p(x,t_{\rm c}\tau + t_{\rm h})$  and the quantities in the numerator are time-rescaled probability currents $J_{\rm I}(x,\tau) = t_{\rm h} j(x,t_{\rm h}\tau)$ and $J_{\rm I}(x,\tau) = t_{\rm c} j(x,t_{\rm c}\tau + t_{\rm h})$. They are interconnected by the continuity equation $\partial P(x,\tau)/\partial \tau = - \partial J(x,\tau)/\partial x$,
which is obtained after inserting the individual expressions into the Fokker-Planck equation (\ref{eq:Fokker-Planck}).

%%%%%%%%%%%%%%%%%%%%%%%%%%%%%%%%%%%%%%%%%%%%%%%%%%%%%%%%%%%%%%%%%%%%%%%%%%%%%%%%%%%%%%%%%%%%%%%%%%
\subsection{Branches II and IV: adiabats}
\label{subs:IIB}
%%%%%%%%%%%%%%%%%%%%%%%%%%%%%%%%%%%%%%%%%%%%%%%%%%%%%%%%%%%%%%%%%%%%%%%%%%%%%%%%%%%%%%%%%%%%%%%%%%

During the adiabats the potential is changed and the hot and cold reservoirs are switched. The adiabaticity of these branches is achieved by performing them so fast that the system state, and thus also its entropy, remains unaltered. The main issue in experimental realization of such branches is to change the reservoir temperature fast enough. This problem was recently solved by Blickle and Bechinger \cite{Blickle2011} within their experimental realization
of a microscopic Stirling cycle. The additional instantaneous change of the potential necessary for realization of the instantaneous adiabatic branches can be performed simply by changing the intensity of the laser employed in the tweezers, which can be done very fast. For a study of a low dissipation model where the adiabatic branches are performed in finite time see e.g. the study \cite{Hu2013} by Hu et al.

During the instantaneous adiabats no heat is obtained from reservoirs and thus the work done by the engine is given just by the decrease of its internal energy,
\begin{equation}
W_{\rm II, \rm IV} = - \Delta E_{\rm II, \rm IV}\,.
\label{eq:adiabatic work}
\end{equation}
Since duration of the adiabatic branches is considered to be negligible, the total duration of one operational cycle of the engine is given by $t_{\rm p} = t_{\rm h} + t_{\rm c}$.

%%%%%%%%%%%%%%%%%%%%%%%%%%%%%%%%%%%%%%%%%%%%%%%%%%%%%%%%%%%%%%%%%%%%%%%%%%%%%%%%%%%%%%%%%%%%%%%%%%
\subsection{Useful and dud heat engines}
\label{subs:IIC}
%%%%%%%%%%%%%%%%%%%%%%%%%%%%%%%%%%%%%%%%%%%%%%%%%%%%%%%%%%%%%%%%%%%%%%%%%%%%%%%%%%%%%%%%%%%%%%%%%%

The work done by the engine per cycle is given by the sum of works done during the individual branches: $W_{\rm out} = W_{\rm I} + W_{\rm II} + W_{\rm III} + W_{\rm IV}$. The mean energy of the system is a state function and hence $\Delta E_{\rm I} + \Delta E_{\rm II} + \Delta E_{\rm III} + \Delta E_{\rm IV} = 0$. Therefore the output work can be written as
\begin{equation}
W_{\rm out} = (T_{\rm h} - T_{\rm c})\Delta S - W^{\rm irr}_{\rm I} - W^{\rm irr}_{\rm III}\,.
\label{eq:w_cycle2}
\end{equation}
It is determined \emph{solely} by the increase of the system entropy during the hot isotherm (\ref{eq:system_entropy}) and by the two irreversible works (\ref{eq:irreversible_work}). 

The upper bound on the output work (\ref{eq:w_cycle2}) is given by the value $W_{\rm out}^{\rm Eq} = (T_{\rm h} - T_{\rm c})\Delta S$ obtained for the equilibrium Carnot cycle and thus it can be attained only when the 
PDF during the whole cycle equals to the equilibrium one.  The PDF is continuous and thus the equilibrium distribution before and after the adiabatic branches must be the same. This is possible only when the sudden changes of temperature during the adiabatic branches are suitably balanced by the discontinuities of the driving \cite{Sekimoto2000, Holubec2014}. 

In order to illustrate general findings given in this paper we review in Tab.~\ref{tab:potential_state}
the potentials for which analytical solutions of the Fokker-Planck equation~(\ref{eq:Fokker-Planck}) are known. From these examples, the Carnot efficiency can be achieved only by engines based on the potentials with the time-dependent stiffness $k(t)$. Namely by the breathing parabola (the third line in Tab.~\ref{tab:potential_state}) and by the log-harmonic potential (the fourth line in Tab.~\ref{tab:potential_state}).
On the other hand, the ``engine'' based on the sliding parabola with fixed $k$ (the second line in Tab.~\ref{tab:potential_state}) cannot operate without a power supply. According to Tab.~\ref{tab:potential_state} and Eq.~(\ref{eq:w_cycle2}) the work produced by this ``engine'' is never positive, $W_{\rm out} = -\int_{0}^{t_{\rm p}}{d}t\,\frac{1}{\mu(t)}\left[\dot{m}(t)\right]^2 \le 0 < W_{\rm out}^{\rm Eq}$, i.e.,  most of the energy transferred to the system is dissipated and the maximum achievable work is zero. The value is obtained whenever the time derivative of the mean particle position, $\dot{m}(t) = {d}m/{d}t$, vanishes, i.e., in case the potential is not varied at all. 

The question arises why the potentials with time-dependent stiffness $k(t)$ may support a useful heat engine, while the potential with only time-dependent position of the minimum $y(t)$ can not. The answer is quite simple. In the two cases with time-dependent stiffness the output works are proportional to the variances $\sigma(t)$ and $4(\nu+1)f(t)$ of the corresponding PDFs which both depend on reservoir temperatures.
On the other hand, in case of the sliding parabola, $W_{\rm out}$ is proportional to the mean position of the particle $m(t)$, which is temperature independent \cite{Schmiedl2007}. The output work is unaffected by the presence of the heat reservoirs and thus this ``engine'' indeed can not transform heat into a useful work.

%%%%%%%%%%%%%%%%%%%%%%%%%%%%%%%%%%%%%%%%%%%%%%%%%%%%%%%%%%%%%%%%%%%%%%%%%%%%%%%%%%%%%%%%%%%%%%%%%%%%%%%%%%%%%%%%%%
\section{Power optimization procedure}
%%%%%%%%%%%%%%%%%%%%%%%%%%%%%%%%%%%%%%%%%%%%%%%%%%%%%%%%%%%%%%%%%%%%%%%%%%%%%%%%%%%%%%%%%%%%%%%%%%%%%%%%%%%%%%%%%%
\label{sec:max_eta}

Most common performance characteristics of heat engines are their efficiency $\eta$ and output power $P$. The larger these two variables are, the better engine we have. 
Traditionally \cite{Seifert2012, Schmiedl2008,Verley2014} they are defined as $\eta = W_{\rm out}/Q_{\rm h}$ and $P=W_{\rm out}/t_{\rm p}$, where $Q_{\rm h} = \Delta E_{\rm I} + W_{\rm I}$ is the amount of heat accepted by the engine from the hot reservoir.

Here we focus on the power-maximizing procedure proposed in \cite{Schmiedl2008, Muratore-Ginanneschi2015}. It proceeds in two steps. First, the durations of the isothermal branches and also the PDFs at their ends, $p(x,0)$ and $p(x,t_{\rm h})$, are fixed and the irreversible works $W^{\rm irr}_{\rm I}$ and $W^{\rm irr}_{\rm III}$ are minimized. This step yields the optimal form of the potential $U(x,t)$ compatible with the given functions $p(x,0)$ and $p(x,t_{\rm h})$. Second, for the optimal potential found in the first step, the power is further optimized as a function of the durations of the two isotherms $t_{\rm h}$ and $t_{\rm c}$. In this section we will focus on the first part of the optimization procedure only. Its second part will be discussed in Sec.~\ref{sec:maxP}.

Using Eq.~(\ref{eq:isothermal_work}) the formula for the input heat $Q_{\rm h} = \Delta E_{\rm I} + W_{\rm I}$ can be written as $Q_{\rm h} = - W^{\rm irr}_{\rm I} + T_{\rm h}\Delta S$. Expressing $\Delta S$ from (\ref{eq:w_cycle2}) and inserting the result into the former expression we obtain
$\eta_{\rm C} Q_{\rm h} = W_{\rm out} + T_{\rm c}/T_{\rm h}W^{\rm irr}_{\rm I} + W^{\rm irr}_{\rm III}$. This formula can be further simplified. the result is
\begin{equation}
Q_{\rm h} = \left[ W_{\rm out} + T_{\rm c}\Delta S_{\rm tot} \right]/\eta_{\rm C}\,,
\label{eq:Qh}
\end{equation}
where
\begin{equation}
\Delta S_{\rm tot} = \frac{1}{T_{\rm h}} W^{\rm irr}_{\rm I}  +  \frac{1}{T_{\rm c}} W^{\rm irr}_{\rm III} \ge 0
\label{eq:total_entropy}
\end{equation}
is the increase of the entropy of the universe per cycle. Using Eqs.~(\ref{eq:w_cycle2})-(\ref{eq:Qh}) the performance can be expressed in terms of $\Delta S$ and the irreversible works $W^{\rm irr}_{\rm I, III}$: 
\begin{eqnarray}
P_{\rm out} &=& \frac{W_{\rm out}}{t_{\rm p}} = \frac{(T_{\rm h} - T_{\rm c})\Delta S - W^{\rm irr}_{\rm I} - W^{\rm irr}_{\rm III}}{t_{\rm p}}\,,
\label{eq:power}\\
\eta &=& \frac{W_{\rm out}}{Q_{\rm h}} = \frac{\eta_{\rm C}}{1 + T_{\rm c}\Delta S_{\rm tot}/W_{\rm out}}\,.
\label{eq:efficiency}
\end{eqnarray}
For $W_{\rm out} > 0$ the positivity of $\Delta S_{\rm tot}$ implies $\eta \le \eta_{\rm C}$.

During the first step of the optimization procedure the irreversible works $W^{\rm irr}_{\rm I, III}$ are minimized for fixed $t_{\rm h}$, $t_{\rm c}$, $p(x,0)$ and $p(x,t_{\rm h})$ and thus also for fixed $\Delta S$, see Eq.~(\ref{eq:system_entropy}). For fixed PDFs $p(x,0)$ and $p(x,t_{\rm h})$ the functionals $W_{\rm I}^{\rm irr}$ and $W_{\rm III}^{\rm irr}$ are independent.
Under these conditions the optimal potential optimizes at the same time the total entropy production per cycle $\Delta S_{\rm tot}$ (minimum), the output work $W_{\rm out}$ (maximum) and, in case $W_{\rm out} > 0$, also the efficiency (maximum). Moreover, if one does not struggle to find the optimal potential but rather optimizes $W^{\rm irr}_{\rm I, \rm III}$ within a restricted class of potentials compatible with given $p(x,0)$ and $p(x,t_{\rm h})$, this observation still holds: within the restricted class of potentials and for given parameters, the suboptimal potentials yield maximum efficiency and output power and minimize $\Delta S_{\rm tot}$. However, one should have in mind that this result is valid only under rather strong restriction of fixed $t_{\rm h}$, $t_{\rm c}$, $p(x,0)$ and $p(x,t_{\rm h})$. 

Specific forms of potentials which optimize the individual irreversible works shown in Tab.~\ref{tab:potential_state} can be obtained from the corresponding response functions $m(t)$, $\sigma(t)$ and $f(t)$. These are given by \cite{Schmiedl2007, Holubec2014a}
\begin{align}
m(t) &= c_1(1+c_2 t)\,,\label{eq:mopt}\\
\sigma(t) &= c_1(1+c_2 t)^2\,,\label{eq:sopt}\\
f(t) &= c_1(1+c_2 t)^2\label{eq:fopt}\,.
\end{align}
Note that the brackets are raised to powers which correspond to the order of the individual moments: $m(t) = \left<x\right>$, $\sigma(t) = 4(\nu+1)f(t) = \left< [x-m(t)]^2\right>$. The constants $c_1$ and $c_2$ for the individual isothermal branches differ and must be determined from the fixed distributions $p(x,0)$ and $p(x,t_{\rm h})$.  

In order to demonstrate these findings we consider the log-harmonic potential (last line in Tab.~\ref{tab:potential_state})
\begin{equation}
U(x,t) = -(2\nu+1)T(t)\ln|x|+\frac{1}{2}k(t)x^2\,.
\label{eq:UlogHarm}
\end{equation}
For this potential the PDF assumes the form $p(x,t)=\frac{[f(t)]^{-\nu-1}}{\Gamma(\nu+1)}\left(\frac{x}{2}\right)^{2\nu+1}\exp\left[-\frac{x^2}{4f(t)}\right]$ and thus it is solely determined by the variance $4(\nu+1)f(t)$. Hence in order to specify the distributions $p(x,0)$ and $p(x,t_{\rm h})$ it is enough to fix the variances at the ends of the two isothermal branches, i.e., to fix $f(0)=f(t_{\rm p})=f_0$ and $f(t_{\rm h}) = f_1$. 

%%%%%%%%%%%%%%%%%%%%%%%%%%%%%%%%%%%%%%%%%%%%%%%%%%%%%%%%%%%%%%%%%%
\begin{figure}
	\centering
		\includegraphics[width=1.0\columnwidth]{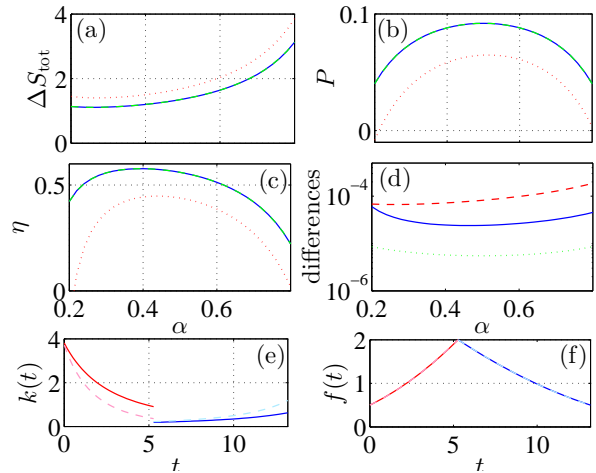}
	\caption{(Color
  online) Panels (a)-(c): performance of the engine which exhibits the optimal response ($n=2$ in (\ref{eq:f_max_power}), solid blue lines), the optimized response for $n=3$ (green dashed lines) and a non-optimized response for $n=3$ (dotted red lines) as the functions of $\alpha = t_{\rm h}/t_{\rm p}$. 
In these panels the green dashed and solid blue lines overlap. 	Panel (d) shows the differences between the optimal ($n=2$) and suboptimal ($n=3$) quantities. The dotted green line shows the difference between the optimal and the suboptimal power, the solid blue line shows the same for efficiency and the dashed red line depicts the difference between the suboptimal and the optimal entropy production. Panel (e): the optimal (solid line) and the suboptimal (dashed line) driving for $\alpha = 2/5$. Panel (f) shows the corresponding response functions $f(t)$. Parameters taken: $T_{\rm h} = 4$, $T_{\rm c} = 0.5$, $\mu_{\rm h} = \mu_{\rm c} = 1$, $f_0 = 0.5$, $f_1 = 2$, $\nu = 1$ and $t_{\rm p} \approx 13.19$, $t_{\rm p} = t_{\rm p}^{\star}$, see Eq.~(\ref{eq:tp_opt}).}
	\label{fig:sub_optimal}
\end{figure}
%%%%%%%%%%%%%%%%%%%%%%%%%%%%%%%%%%%%%%%%%%%%%%%%%%%%%%%%%%%%%%%%%%

\begin{figure}
	\centering
		\includegraphics[width=1.0\columnwidth]{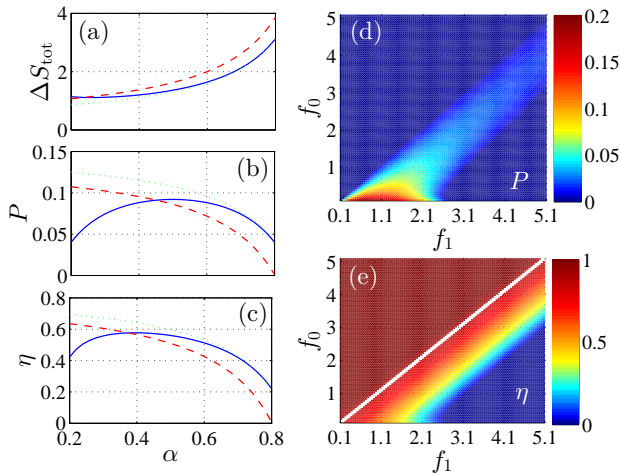}
	\caption{(Color
  online) Panels (a)-(c): performance of the engine for the optimal response with $t_{\rm h} = \alpha t_{\rm p}$ (blue solid line), optimal response with $t_{\rm h} = 10$ (green dotted line) and non-optimal response ($n=3$) with $t_{\rm h} = 10$ (red dashed line) as functions of $\alpha = t_{\rm h}/t_{\rm p}$. Panels (d) and (e) shows the power and efficiency for the optimal response (\ref{eq:f_max_power}) with $t_{\rm h} = 10$ as functions of fixed $f_0$ and $f_1$. Remaining parameters coincide with those in Fig.~\ref{fig:sub_optimal}.}
	\label{fig:super_optimal}
\end{figure}

Imagine that the optimal $f(t)$ (\ref{eq:fopt}) for this potential cannot be obtained. In order to follow the optimization procedure one can then fix the parameters $t_{\rm h}$, $t_{\rm c}$, $T_{\rm h}$, $T_{\rm c}$, $\mu_{\rm h}$, $\mu_{\rm c}$, $\nu$, $f_0$ and $f_1$ and minimize the irreversible works $W^{\rm irr}_{\rm I, \rm III}$ numerically within a certain class of response functions. We consider the class
\begin{equation}
 f(t)=\bigg\{\begin{array}{l l}
    f_0\left(1+a_1t+b_1t^2\right)^n\,,&  t \in [0,t_{\rm h}]\\
    f_1\left[1+a_2(t - t_{\rm h})+b_2(t - t_{\rm h})^2\right]^n\,, &  t \in [t_{\rm h},t_{\rm p}]
  \end{array}\,,
\label{eq:f_max_power}
\end{equation}
parametrized by the real parameters $a_1$ and $a_2$ and by the exponent $n$. The constants $b_1=[(f_1/f_0)^{1/n}-1-a_1 t_{\rm h}]/t_{\rm h}^2$ and $b_2=[(f_0/f_1)^{1/n}-1-a_2 t_{\rm c}]/t_{\rm c}^2$ are fixed by the boundary conditions $f(t_{\rm h}) = f_1$ and $f(t_{\rm p}) = f_0$ respectively. Note that the class (\ref{eq:f_max_power}) contains also the optimal response (\ref{eq:fopt}), which follows for $n=2$, $a_1 = \left(\sqrt{f_1/f_0} - 1\right)/t_{\rm h}$, $a_2 = \left(\sqrt{f_0/f_1} - 1\right)/t_{\rm c}$ and $b_1=b_2=0$ (or for $n=1$ and different $a_{1,2}$, $b_{1,2}$). The driving $k(t)$ corresponding to the individual responses $f(t)$ (and thus also the specific form of the correspond-
ing potential) can be calculated from \cite{Holubec2014}
\begin{equation}
k(t) = \frac{\mu(t) T(t) - \dot{f}(t)}{2 \mu(t) f(t)}\,.
\label{eq:f-k}
\end{equation}
For the potential (\ref{eq:UlogHarm}) the irreversible works (\ref{eq:irreversible_work}) are given by \cite{Holubec2014}
$W^{\rm irr}_{\rm I} = \frac{\nu+1}{\mu_{\rm h}}\int_0^{t_{\rm h}}dt\, \frac{[\dot{f}(t)]^2}{f(t)}$ and $W^{\rm irr}_{\rm III} = \frac{\nu+1}{\mu_{\rm c}}\int_{t_{\rm h}}^{t_{\rm p}}dt\, \frac{[\dot{f}(t)]^2}{f(t)}\,$.
We have numerically minimized these expressions as functions of $a_1$ and $a_2$ for the response functions (\ref{eq:f_max_power}) with $n = -1$, $1/2$, $1$, $2$ and $3$.
From these $n$ values, the optimization procedure gave reasonable results only in cases $n= 1$, $2$ and $3$. For $n = -1$, $1/2$ the resulting values of $a_1$ and $a_2$ were either complex, or the corresponding driving $k(t)$ assumed negative values which is in our setting unphysical ($k(t)$ stands for the stiffness of a spring). Further, as mentioned above, the optimal responses for $n=1$ and $n=2$ both coincide
with the optimal response (\ref{eq:fopt}). Therefore, in Figs.~\ref{fig:sub_optimal} and \ref{fig:super_optimal}, we show only the 
results for the optimal case ($n = 2$) and for the case $n=3$.

Fig.~\ref{fig:sub_optimal}(a-c) shows the total entropy production (\ref{eq:total_entropy}), output power (\ref{eq:power}) and efficiency (\ref{eq:efficiency}) as functions of the allocation of the cycle period $t_{\rm p}$ among the two isothermal branches: 
\begin{equation}
\alpha = \frac{t_{\rm h}}{t_{\rm p}}\,,\qquad \alpha \in [0,1]\,.
\label{eq:alpha}
\end{equation}
The curves calculated for the optimal response (\ref{eq:fopt}) are solid blue, dashed green curves correspond to the optimized response (\ref{eq:f_max_power}) with $n=3$. The dotted red line in panels (a)-(c) show the performance of the engine which exhibits the response (\ref{eq:f_max_power}) for $n = 3$ and the non-optimized parameters $a_1 = -0.1 a_1^{\star}$ and $a_2 = -0.1 a_2^{\star}$, where $\star$ denotes the corresponding optimized ones. Clearly the performance is worse than that for the optimized protocols. 

%%%%%%%%%%%%%%%%%%%%%%%%%%%%%%%%%%%%%%%%%%%%%%%%%%%%%%%%%%%%%%%%%%
%%%%%%%%%%%%%%%%%%%%%%%%%%%%%%%%%%%%%%%%%%%%%%%%%%%%%%%%%%%%%%%%%%

It is remarkable that the optimal response (solid lines) and the suboptimal one (dashed lines) plotted in Fig.~\ref{fig:sub_optimal}(f) are nearly indistinguishable, although the corresponding drivings $k(t)$ significantly differ (Fig.~\ref{fig:sub_optimal}(e)). Also these two protocols lead to almost the same engine performance (Fig.~\ref{fig:sub_optimal}(a-c)). The differences between the fully optimized $\Delta S_{\rm tot}$, $P$ and $\eta$ and those optimized only within the class of response functions (\ref{eq:f_max_power}) with $n=3$ are shown in panel (d). The dotted green line shows the difference between the optimal and the suboptimal power, the solid blue line shows the same for efficiency and the dashed red line depicts the difference between the suboptimal and the optimal entropy production. Indeed, the optimal protocol provides better values, but the functions are surprisingly close to each other. 

Although a general procedure for calculation of optimal protocols was described recently \cite{Muratore-Ginanneschi2015}, this result suggests that in cases when it is hard to solve the respective formulas it may be sufficient to use a suboptimal driving/response obtained by numerical minimization of irreversible works within a suitable class of protocols/responses.
For a different way to find a suboptimal protocol numerically we refer to \cite{Then2008}.

Fig.~\ref{fig:super_optimal}(a-c) shows that the performance obtained by the optimal protocol (blue solid lines) is indeed optimal only for the given parameters. The dashed red curve is calculated for the engine which exhibits the response (\ref{eq:f_max_power}) for $n = 3$, non-optimized parameters $a_1 = -0.1 a_1^{\star}$ and $a_2 = -0.1 a_2^{\star}$ and further for fixed duration of the hot isotherm $t_{\rm h} = 10$. This engine may indeed produce more power with higher efficiency than that driven by the optimal protocol with different $t_{\rm h} = \alpha t_{\rm p}$. However, if we calculate the optimal response using $t_{\rm h} = 10$ (dotted green line) we see that it again overcomes the non-optimal results. 

Fig.~\ref{fig:super_optimal}(d-e) shows the power $P$ and the efficiency $\eta$ for the optimal response (\ref{eq:f_max_power}) as functions of the variances at the ends of the isothermal branches, i.e., of $f_0$ and $f_1$. In the region where the engine produces positive work ($P>0$), the power increases with decreasing $f_0$ and thus the cycles where the particle is at the beginning of the hot isotherm as localized as possible should be favored. In the figure the maximum power $P \approx 0.19$ (cf. values of power in Fig.~\ref{fig:universal1}) is attained for the minimal used $f_0 = 0.1$ and for the moderate $f_1 = 0.8$. The corresponding efficiency $\eta \approx 0.65$ is relatively large (for parameters used $\eta_{\rm C} = 0.875$). The efficiency in Fig.~\ref{fig:super_optimal}(e) grows monotonically with $f_0\to f_1$ (the process is more and more quasi-static), while in the limit $f_1 = f_0$ the driving becomes time-independent and the power vanishes.

%%%%%%%%%%%%%%%%%%%%%%%%%%%%%%%%%%%%%%%%%%%%%%%%%%%%%%%%%%%%%%%%%%%%%%%%%%%%%%%%%%%%%%%%%%%%%%%%%%%%%%%%%%%%%%%%
\section{Efficiency at maximum power}
%%%%%%%%%%%%%%%%%%%%%%%%%%%%%%%%%%%%%%%%%%%%%%%%%%%%%%%%%%%%%%%%%%%%%%%%%%%%%%%%%%%%%%%%%%%%%%%%%%%%%%%%%%%%%%%%%%
\label{sec:maxP}

In this section we focus on the second part of the power optimization procedure, i.e., for the optimal irreversible works $W^{\rm irr}_{\rm I, \rm III}$ obtained in the first part of the procedure we maximize the power as a function of the durations of the two isotherms $t_{\rm h}$ and $t_{\rm c}$.

As can be inspected from Eqs.~(\ref{eq:w_cycle2}) and (\ref{eq:power}), in order to determine how the power depends on the times $t_{\rm h}$ and $t_{\rm c}$ this dependence must be first revealed for $W^{\rm irr}_{\rm I, \rm III}$.
While this can be done separately for specific models (see for example Tab.~\ref{tab:potential_state}), authors of the studies \cite{Schmiedl2008, Muratore-Ginanneschi2015} suggested that, after the optimization, the irreversible works (\ref{eq:irreversible_work}) assume the low-dissipation form
\begin{equation}
W^{\rm irr}_{\rm I, \rm III} = \frac{A_{\rm I, \rm III}}{t_{\rm I, \rm III}}\,,
\label{eq:low_diss}
\end{equation}
where the parameters $A_{\rm I}$ and $A_{\rm III}$ depend on the mobilities and the fixed distributions $p(x,0)$ and $p(x,t_{\rm h})$, but not on the times $t_{\rm h}$ and $t_{\rm c}$.
This formula agrees with the optimized irreversible works given in Tab.~\ref{tab:potential_state} for the protocols where the prefactor before the quadratic part of the potential is varied (lines three and four). For the sliding parabola the terms $g_{\rm I, \rm III}$ are $t_{\rm I, \rm III}$ independent and thus the irreversible works remain positive even for infinitely slow driving since the reversible limit does not exist in
this case \cite{Esposito2010b}. This suggests that the formula (\ref{eq:low_diss}) should be applied with care. 
In general, minimization of the functional (\ref{eq:irreversible_work}) for fixed distributions $p(x,0)$ and $p(x,t_{\rm h})$
implies the simple time dependence (\ref{eq:low_diss}) only in case no special form of the potential $U(x,t)$ is assumed a priori.  

In the rest of this paper we will assume that the relation (\ref{eq:low_diss}) holds, i.e., we will investigate performance of a general low-dissipation heat engine. The diffusion-based heat engines described in Secs.~\ref{sec:model}-\ref{sec:max_eta} will serve us only for illustrations. In the rest of this section we will review the results on the EMP for low-dissipation heat engines obtained in the studies \cite{Schmiedl2008, Esposito2010b}. These results will be utilized in Sec.~\ref{sec:tradeoff} for investigation of optimal trade-off between power and efficiency.

If we express $t_{\rm h}$ and $t_{\rm c}$ using the whole duration of the cycle as $t_{\rm h} = \alpha t_{\rm p}$ and $t_{\rm c} = (1-\alpha) t_{\rm p}$ the output power (\ref{eq:power}) reads
\begin{equation}
P = \frac{(T_{\rm h} - T_{\rm c})\Delta S}{t_{\rm p}} - \frac{(1-\alpha)A_{\rm I} +  \alpha A_{\rm III}}{t_{\rm p}^2\alpha(1-\alpha)}\,.
\label{eq:power_Wirr}
\end{equation} 
Maximizing the power as the function of $t_{\rm p}$ we obtain the optimal cycle duration with respect to the allocation $\alpha$ of the cycle period $t_{\rm p}$ between the two isotherms:
\begin{equation}
t_{\rm p}^{\alpha} = \frac{2}{T_{\rm h}\eta_{\rm C} \Delta S}\frac{(1-\alpha)A_{\rm I} + \alpha A_{\rm III}}{\alpha(1-\alpha) }\,,
\label{eq:tp_alpha}
\end{equation}
The corresponding power (\ref{eq:power_Wirr}) and efficiency (\ref{eq:efficiency}) read
\begin{eqnarray}
P^{\alpha} &=& \frac{1}{4}\frac{ (1 - \alpha)\alpha}{ (1 - \alpha)A_{\rm I} +
     \alpha A_{\rm III}}\left(T_{\rm h} \eta_{\rm C} \Delta S\right)^2\,,
\label{eq:P_alpha}\\
\eta^{\alpha} &=& \frac{ (1 - \alpha)A_{\rm I} +
     \alpha A_{\rm III}}{
  (2-\eta_{\rm C}) (1 - \alpha)A_{\rm I} + 2  \alpha A_{\rm III}}\eta_{\rm C}\,.
\label{eq:eta_alpha}
\end{eqnarray}
Interestingly enough, the resulting optimal output work (\ref{eq:w_cycle2}) is given by $W_{\rm out}^{\alpha} = T_{\rm h}\eta_{\rm C}\Delta S/2 = W_{\rm out}^{\rm Eq}/2$ and, therefore, is determined by the temperatures and by the fixed distributions $p(x,0)$ and $p(x,t_{\rm h})$ only. The power (\ref{eq:P_alpha}) exhibits a maximum as the function of $\alpha$, the efficiency (\ref{eq:eta_alpha}) monotonically decreases. Its maximum value $\eta^{\alpha} = \eta_{\rm C}/(2-\eta_{\rm C})$ is attained at $\alpha = 0$ (or equivalently for $A_{\rm III}/A_{\rm I} \to 0$). This efficiency is smaller than $\eta_{\rm C}$ (for $\alpha = 0$ the cold isotherm becomes reversible, but the hot one runs infinitely fast). Minimum value of $\eta^{\alpha}$ attained at $\alpha = 1$ (or equivalently for $A_{\rm III}/A_{\rm I} \to \infty$) is $\eta_{\rm C}/2$. 
The efficiency at maximum power is thus bounded by the inequalities $\eta_{\rm C}/2 \le \eta^{\alpha} \le \eta_{\rm C}/(2-\eta_{\rm C})$ \cite{Esposito2010b}. For the both limiting values of $\alpha$ the optimal cycle duration $t_{\rm p}^{\alpha}$ diverges and thus the corresponding power vanishes.

%%%%%%%%%%%%%%%%%%%%%%%%%%%%%%%%%%%%%%%%%%%%%%%%%
\begin{figure}
	\centering
		\includegraphics[width=1.0\columnwidth]{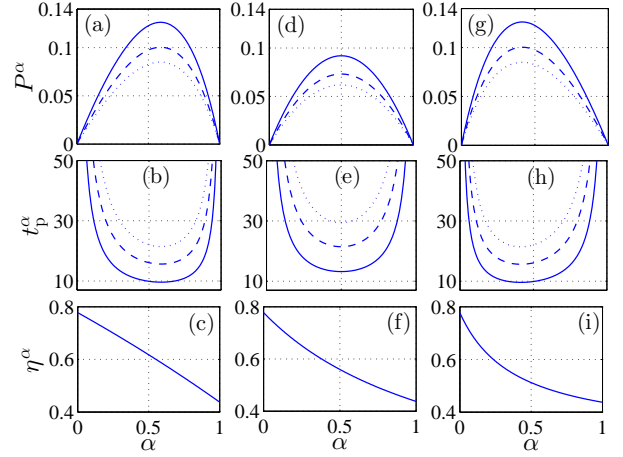}
	\caption{(Color
  online) Performance of the engine working at the optimal power $P^{\alpha}$ (\ref{eq:P_alpha}) as the function of the allocation of the cycle period $t_{\rm p}$ among the two isothermal branches $\alpha$ (\ref{eq:alpha}). All quantities are plotted for three values of the variance at the end of the hot isotherm, $4(\nu+1)f_1$: $f_1 = 4$ (dotted lines), $f_1 =3$ (dashed lines) and $f_1 =2$ (solid lines). Panel (a)  shows the optimal power (\ref{eq:P_alpha}) for $2\mu_{\rm h} = \mu_{\rm c} = 2$. The corresponding  cycle duration (\ref{eq:tp_alpha}) is depicted in panel (b) and the efficiency (\ref{eq:eta_alpha}) is plotted in panel (c).
Panels (d)-(f) and (g)-(i) show the same for the mobilities $\mu_{\rm h} = \mu_{\rm c} = 1$ and $\mu_{\rm h} = 2\mu_{\rm c} = 2$ respectively. Other parameters used: $T_{\rm h} = 4$, $T_{\rm c} = 0.5$, $f_0 = 0.5$, $\nu = 1$ and $t_{\rm p} \approx 13.19$, $t_{\rm p} = t_{\rm p}^{\star}$, see Eq.~(\ref{eq:tp_opt}).}
	\label{fig:universal1}
\end{figure}
%%%%%%%%%%%%%%%%%%%%%%%%%%%%%%%%%%%%%%%%%%%%%%%%%

Optimization of the output power $P^{\alpha}$ with respect to $\alpha$ yields the optimal time redistribution 
\begin{equation}
\alpha^{\star} = \frac{A_{\rm I} - \sqrt{A_{\rm I}A_{\rm III}}}{A_{\rm I} - A_{\rm III}}.
\label{eq:alpha_opt}
\end{equation}
Notice that, $\alpha^{\star}$ not only maximizes $P^{\alpha}$, but also minimizes the optimal cycle duration $t_{\rm p}^{\alpha}$. This is because the power (\ref{eq:P_alpha}) can be rewritten as $P^{\alpha} = T_{\rm h} \eta_{\rm C} \Delta S/(2 t_{\rm p}^{\alpha})$. Also note that in the symmetric case ($A_{\rm I} = A_{\rm III}$) we have $\alpha^{\star} = 1/2$ \cite{Schmiedl2008}. 

The duration of the optimal cycle $t_p^{\star}$, the optimal power, $P^{\star}$ and the
corresponding efficiency $\eta^{\star}$ can be obtained by substituting $\alpha^{\star}$ for $\alpha$
in Eqs.~(\ref{eq:tp_alpha}), (\ref{eq:P_alpha}) and (\ref{eq:eta_alpha}) respectively. The results are \cite{Schmiedl2008}
\begin{eqnarray}
t_{\rm p}^{\star} &=& \frac{2}{T_{\rm h}\eta_{\rm C}\Delta S}(\sqrt{A_{\rm I}}+\sqrt{A_{\rm III}})^2\,,
\label{eq:tp_opt}\\
P^{\star} &=& \frac{1}{4}\left(\frac{T_{\rm h}\eta_{\rm C}\Delta S}{\sqrt{A_{\rm I}} + \sqrt{A_{\rm III}}}\right)^2\,,
\label{eq:P_opt}\\
\eta^{\star} &=& \frac{\eta_{\rm C}(1+\sqrt{A_{\rm III}/A_{\rm I}})}{2(1+\sqrt{A_{\rm III}/A_{\rm I}})  - \eta_{\rm C}}
\label{eq:eta_opt}\,.
\end{eqnarray}
Note that the EMP (\ref{eq:eta_opt}) does not depend on the individual parameters $A_{\rm I}$ and $A_{\rm III}$, but only on their ratio $A_{\rm I}/A_{\rm III}$. For the diffusion-based heat engines described in Secs.~\ref{sec:model}-\ref{sec:max_eta} this ratio is determined by the mobilities, $A_{\rm I}/A_{\rm III} = \mu_{\rm c}/\mu_{\rm h}$. Hence the final result (\ref{eq:eta_opt}) for the EMP is independent not only of the distributions $p(x,0)$ and $p(x,t_{\rm h})$, but also of the used potential $U(x,t)$ \cite{Schmiedl2008}.

Fig.~\ref{fig:universal1} shows the optimal power $P^{\alpha}$ (upper panels), the corresponding cycle duration $t^{\alpha}$ (middle panels) and efficiency $\eta^{\alpha}$ (lower panels) for the engine based on the log-harmonic potential (\ref{eq:UlogHarm}) (corresponding parameters $A_{{\rm I}, {\rm III}}$ can be found in Tab.~\ref{tab:potential_state}). With increasing variance of the particle distribution at the end of the hot isotherm, $4(\nu+1)f_1$, the optimal power decreases [panels (a), (d), (g)] and the duration of the optimal cycle increases [panels (b), (e), (h)]. This happens because the parameters $A_{{\rm I}, {\rm III}} \propto (\sqrt{f1}-\sqrt{f0})^2$ increases with $f_1$ much faster then the system entropy during the hot isotherm (\ref{eq:system_entropy}), $\Delta S \propto \ln f_1/f_0$. Panels (c), (f) and (i) show that the efficiency at maximum power (\ref{eq:eta_opt}) indeed depends solely on temperatures and the ratio of the mobilities $A_{\rm I}/A_{\rm III} = \mu_{\rm c}/\mu_{\rm h}$, not on the fixed distributions $p(x,0)$ and $p(x,t_{\rm h})$. Panels (a)-(c) in Fig.~\ref{fig:universal1} are calculated for the mobilities $2\mu_{\rm h}=\mu_{\rm c}=2$, panels (d)-(f) for $\mu_{\rm h}=\mu_{\rm c}=1$ and panels (g)-(i) for $\mu_{\rm h}=2\mu_{\rm c}=2$. From formulas (\ref{eq:tp_opt})-(\ref{eq:P_opt}) for $t^{\star}$ and $P^{\star}$ we see that these variables are symmetric with respect to $A_{{\rm I}, {\rm III}}$ and thus also with respect to the mobilities $\mu_{\rm h}$ and $\mu_{\rm c}$. Furthermore, the parameters $A_{{\rm I}, {\rm III}}$ in (\ref{eq:tp_opt})-(\ref{eq:P_opt}) increase in inverse proportion to the mobilities and thus larger $\mu_{\rm h}$ and $\mu_{\rm c}$ yield smaller $t^{\star}$ and larger $P^{\star}$. For example consider the curves for $f_1 = 2$ in Fig.~\ref{fig:universal1} (solid lines). Then the optimal power $P^{\star}\approx 0.13$ is attained at $t_{\rm p}^{\star}\approx 9.6$ both for $2\mu_{\rm h}=\mu_{\rm c}=2$ and for $\mu_{\rm h}=2\mu_{\rm c}=2$. For the same $f_1$ and equal mobilities, $\mu_{\rm h}=\mu_{\rm c}=1$, the obtained optimal power is smaller ($P^{\star}\approx 0.09$) and the corresponding cycle duration longer ($t_{\rm p}^{\star}\approx 13.2$). The largest efficiency at maximum power in Fig.~\ref{fig:universal1}, $\eta^{\star} \approx 0.59$, is attained at $2\mu_{\rm h}=1\mu_{\rm c}=2$ (smaller mobility during the hot isotherm). 
The smallest efficiency at maximum power, $\eta^{\star}\approx 0.53$, corresponds to $2\mu_{\rm h}=\mu_{\rm c}=2$ (larger mobility during the hot isotherm). For $\mu_{\rm h}=\mu_{\rm c}=1$ we obtained $\eta^{\star}\approx 0.56$.

%%%%%%%%%%%%%%%%%%%%%%%%%%%%%%%%%%%%%%%%%%%%%%%%%%%%%%%%%%%%%%%%%%%%%%%%%%%%%%%%%%%%%%%%%%%%%%%%%%%%%%%%%%%%%%%%
\section{Efficiency near maximum power}
%%%%%%%%%%%%%%%%%%%%%%%%%%%%%%%%%%%%%%%%%%%%%%%%%%%%%%%%%%%%%%%%%%%%%%%%%%%%%%%%%%%%%%%%%%%%%%%%%%%%%%%%%%%%%%%%%
\label{sec:tradeoff}

In this section we derive our main result, i.e., we show that any trade-off measure expressible in terms of efficiency $\eta$ and the ratio $P/P^\star$ of power and its optimal value can be discussed on the same basis as the EMP (\ref{eq:eta_opt}), i.e., only as a function of the ratio 
\begin{equation}
A = \sqrt{\frac{A_{\rm III}}{A_{\rm I}}} > 0
\label{eq:B}\,.
\end{equation}
This is possible because $A_{\rm I}$, $A_{\rm III}$ and $\Delta S$ in the formulas for power and efficiency behave merely as scaling and shifting constants. To be more specific, it is favorable to study the trade-off with respect to the point of maximum power determined by the optimal cycle duration $t_p^{\star}$ (\ref{eq:tp_opt}) and the optimal time redistribution $\alpha^{\star}$ (\ref{eq:alpha_opt}). Natural choice of the dimensionless coordinates suitable for such description is
\begin{align}
\tau &= \frac{t_{\rm p}}{t_p^{\star}} - 1\,,& \tau &\in [-1,\infty)
\label{eq:tau}\,,\\
a &= \frac{\alpha}{\alpha^{\star}}-1\,,& a &\in [-1,\frac{1}{\alpha^{\star}} - 1]
\label{eq:a}\,.
\end{align}
The point of maximum power corresponds in these coordinates to the origin, i.e., $\tau=a=0$. The parameter $\tau$ is larger than zero whenever $t_{\rm p} > t_p^{\star}$ and similarly for the parameter $a$.  

Let us now insert the formulas (\ref{eq:B})-(\ref{eq:a}) into the equations for power (\ref{eq:power}) and efficiency (\ref{eq:efficiency}) and calculate \emph{the relative loss of power} when one does not use the optimal $t_{\rm p}$ and $\alpha$
\begin{equation}
\delta_P = \frac{P - P^{\star}}{P^{\star}} \le 0\,,
\label{eq:reldP_def}
\end{equation}
and the corresponding \emph{relative change in efficiency}
\begin{equation}
\delta_{\eta} = \frac{\eta - \eta^{\star}}{\eta^{\star}}\,.
\label{eq:reldeta_def}
\end{equation}
The results read
\begin{align}
\delta_P &= \frac{a^2}{(1 + a) (a - A) (1 + \tau)^2} - \left(\frac{\tau}{1 + \tau}\right)^2
\,,\label{eq:reldP}\\
\delta_{\eta} &= - 1 + \frac{2(1+A)-\eta_{\rm C}}{a - A}\times\nonumber\\
&\times\frac{a (2 a - A + 1) - A + 2 (1 + a) (a - A) \tau}{2(1 + \tau)(1 + a)(1 + A) - \eta_{\rm C}}\,.
\label{eq:reldeta}
\end{align}
The relative loss of power $\delta_P$ depends solely on the parameters $\tau$, $a$ and $A$ while the relative change in efficiency $\delta_{\eta}$ additionally contains the Carnot efficiency $\eta_{\rm C}$. The EMP (\ref{eq:eta_opt}) also depends solely on $A$ and $\eta_{\rm C}$, $\eta^{\star} = (1+A)\eta_{\rm C}/(2+2A-\eta_{\rm C})$, and thus Eq.~(\ref{eq:reldeta})
implies that the efficiency itself can be written as a function of $A$ and $\eta_{\rm C}$. To sum up, $A$ is the only model-dependent parameter which enters the expressions for $P/P^{\star}$ and $\eta$. This means that the optimization with respect to $a$ and $\tau$ of any variable which can be expressed in terms of $P/P^{\star}$ and $\eta$ can be performed at the same time for the whole class of low-dissipation heat engines specified by a fixed value of the dimensionless parameter $A$. While all the parameters $A_{\rm I}$, $A_{\rm III}$ and $A$
depend on the specific system in question, only the parameter $A$ can be often independent of most details of system dynamics and of its coupling to thermal reservoirs. Prominent examples are the diffusion-based heat engines, given that the used driving minimizes the work dissipated during the isothermal branches as discussed in Secs.~\ref{sec:model} and \ref{sec:max_eta}. In this case, different potentials lead to different parameters $A_{\rm I}$ and $A_{\rm III}$ (see for example the expressions in Tab.~{\ref{tab:potential_state}}). On the other hand the parameter $A$ is determined for any potential just by the ratio of mobilities, $A=\sqrt{\mu_{\rm h}/\mu_{\rm c}}$. The result of optimization of any variable which can be expressed in terms of $P/P^{\star}$ and $\eta$ thus not depend on the used potential. Similar results can be expected also for other systems than diffusion-based heat engines.
This observation constitutes our main result. 

\begin{figure}
	\centering
		\includegraphics[width=1.0\columnwidth]{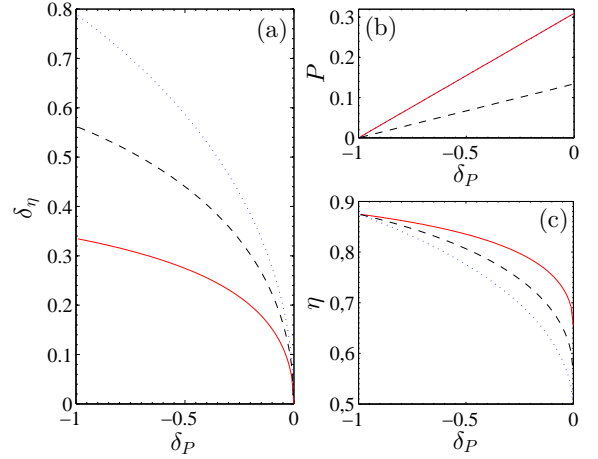}
	\caption{(Color online) Panel (a): the optimal relative gain in efficiency $\delta_{\eta}$ (\ref{eq:reldeta_def}) calculated for fixed values of the loss of power $\delta_{P}$ (\ref{eq:reldP_def}) for $\eta_{\rm C} = 0.875$ and three values of the parameter $A$: $A = \sqrt{0.1}$ (red solid line), $A = 1$ (black dashed line) and $A = \sqrt{10}$ (blue dotted line). The corresponding power and efficiency are shown in panels (b) and (c) respectively. For calculation of the power we have used the parameters $T_{\rm h} = 4$, $T_{\rm c} = 0.5$, $f_0 = 0.5$, $f_1 = 2$ and $\nu = 1$ and further $10\mu_{\rm h} = \mu_{\rm c} = 10$ for $A = \sqrt{0.1}$, $\mu_{\rm h} = \mu_{\rm c} = 1$ for $A = 1$ and $\mu_{\rm h} = 10\mu_{\rm c} = 10$ for $A = \sqrt{10}$.}
	\label{fig:universal4}
\end{figure}

In the rest of this section we present two examples of possible optimization procedures performed on two specific trade-off measures. During the both procedures $A$ is taken as a fixed parameter dictated by the specific system in question. Although our results suggest which values of the parameter $A$ (and also of the parameters $A_{\rm I}$ and $A_{\rm III}$) are preferable from the point of view of large efficiency and large output power ($A_{\rm III} \to 0$ and small, but finite $A_{\rm I}$, see Fig.~\ref{fig:universal4}), the optimization is performed only with respect to the dimensionless coordinates $\tau$ and $a$.

The power exhibits maximum at $\tau=a=0$ and thus $\delta_P$ for small $\tau$ and $a$ varies very slowly. On the other hand, the efficiency can change much more rapidly and thus, for suitable parameters, the loss of power can be much smaller than the gain in efficiency \cite{Chen2001,DeVos1992,Chen1994}. This change in performance can be characterized by the dimensionless trade-off measure 
\begin{equation}
\Gamma(\tau,a) = - 
\frac{\delta_{\eta}}{\delta_{P}} H\!\left(1 + \delta_{P}\right)\,.
\label{eq:tradeoff}
\end{equation}
The function $H$ stands for the Heaviside unit step function. It guarantees that only the parameter region where the system works as a heat engine ($P>0$) is taken into account.  Whenever $\Gamma > 1$ the relative gain in efficiency is larger then the relative loss of power. The natural optimization tasks suitable for this trade-off measure is the following: 
Consider a heat engine which should deliver a certain amount of power $P$, $P<P^{\star}$, in order to operate some machine. Then the value of $\delta_{P}$ is fixed and one should maximize the gain in efficiency $\delta_{\eta}$ for this given $\delta_{P}$, thus also maximizing $\Gamma$.

%%%%%%%%%%%%%%%%%%%%%%%%%%%%%%%%%%%%%%%%%%%%%%%%%%%%%%%%
\begin{figure}
	\centering
		\includegraphics[width=1.0\columnwidth]{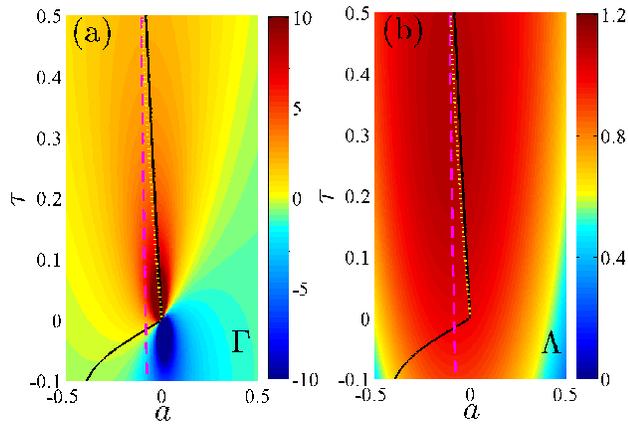}
	\caption{(Color
  online) The trade-off measures $\Gamma$ (\ref{eq:tradeoff}), panel (a), and $\Lambda$ (\ref{eq:Lambda}), panel (b), as functions of the dimensionless parameters $\tau$ and $a$ (\ref{eq:tau})-(\ref{eq:a}) for $\eta_{\rm C} = 0.875$ and $A =  1$. In the both panels the black solid line marks the maximum of $\Gamma$ for fixed $\tau$ and the magenta dashed line highlights the same for $\Lambda$. The yellow dotted lines mark maximum $\Gamma$ for fixed values of $\delta_{\rm P}$ (which is the parameter of these curves).}
	\label{fig:universal2}
\end{figure}
%%%%%%%%%%%%%%%%%%%%%%%%%%%%%%%%%%%%%%%%%%%%%%%%%%%%%%%%

Another function measuring a trade-off between power and efficiency is the simple product of the two, $\eta P$. It exhibits a maximum as a function of $\tau$ and $a$ and a natural optimization task is identification of the respective optimal values. However, the product $\eta P$ depends on the details of the dynamics via the functions $A_{\rm I, III}$ and $\Delta S$ and thus it seems that the optimization should be carried out independently for each set of these functions.
Nevertheless, the rescaled product $\eta P/\eta{^\star}P{^\star}$
depends on the parameter $A$ only. The key observation is that the position of the maximum of the both products is the same. 
This leads us to the second trade-off measure defined as
\begin{equation}
\Lambda(\tau,a) = \left(1 + \delta_{\eta}\right)\left(1 + \delta_{P}\right)H\!\left(1 + \delta_{P}\right)
\,.
\label{eq:Lambda}
\end{equation}

Fig.~\ref{fig:universal2} shows typical shapes of the trade-off measures $\Gamma$ and $\Lambda$ as functions of the dimensionless parameters $a$ and $\tau$. The both functions assume the form of a mountain ridge oriented in the direction of increasing $\tau$ (thus also $t_{\rm p}$). As an eye guide, the maximum heights of the ridge for fixed $\tau$ are highlighted by the black solid line for $\Gamma$ and by the magenta dashed line for $\Lambda$. A general feature of the functions $\Gamma$ and $\Lambda$ is that their maximum values are located in the sector $\tau>0$, $a<0$ of the $\tau$-$a$ diagram. This means that the optimal performance of low-dissipation heat engines according to $\Gamma$ and $\Lambda$ is obtained for $t_{\rm p}>t_{\rm p}^{\star}$ and $\alpha < \alpha^{\star}$.

\begin{figure}
	\centering
		\includegraphics[width=1.0\columnwidth]{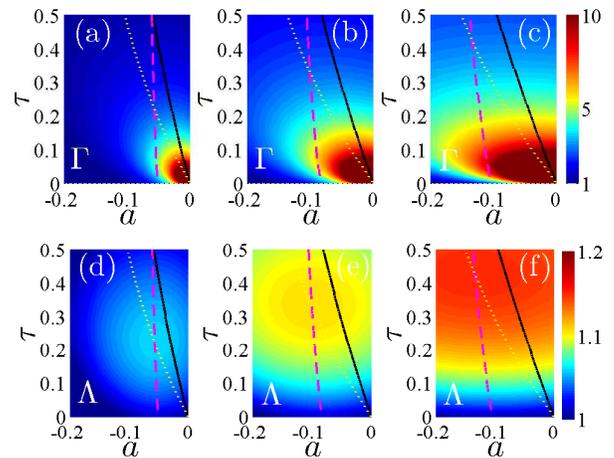}
	\caption{(Color
  online) The trade-off measure $\Gamma$ (\ref{eq:tradeoff}) as the function of the dimensionless parameters $a$ and $\tau$ (\ref{eq:tau})-(\ref{eq:a}) for $\eta_{\rm C} = 0.875$ and for $A = \sqrt{0.1}$ (a), $A = 1$ (b) and $A = \sqrt{10}$ (c). Panels (d)-(f) show the same for the trade-off measure $\Lambda$ (\ref{eq:Lambda}). In all the panels the lines have the same meaning as in Fig.~\ref{fig:universal2}.}
	\label{fig:universal3}
\end{figure}

Fig.~\ref{fig:universal3} depicts how the trade-off measures $\Gamma$ and $\Lambda$ change with the parameter $A$ (\ref{eq:B}). In the figure, $A$ increases from left to right. The panels (a)-(c) show that the region where $\Gamma$ is large grows with increasing $A$. At the same time the maximum value of $\Lambda$ shifts towards lower $a$ and larger $\tau$ values. Note that also values of $\Lambda$ increase with $A$ and thus it can be tempting to conclude at this point that engines characterized by large $A$ values exhibit better performance. However, Figs.~\ref{fig:super_optimal} and \ref{fig:universal4} suggest that the opposite is true. This conflict is just apparent, because the functions $\Gamma$ and $\Lambda$ measure
different aspects of the engine performance than the efficiency. The function $\Gamma$ measures how 
much we loose in power and how much we gain in efficiency if the engine does not work at maximum power and the parameter $A$ is fixed. It says nothing about actual values of efficiency and power. At the same time the important feature of the function $\Lambda$ is that it has the maximum at the same point ($\tau$, $a$) as the function $\eta P$. The specific values of $\Lambda$ for different $A$ values can be strongly affected by the denominator $\eta^{\star} P^{\star}$.

%%%%%%%%%%%%%%%%%%%%%%%%%%%%%%%%%%%%%%%%%%%%%%%%%%%%%%%%
%%%%%%%%%%%%%%%%%%%%%%%%%%%%%%%%%%%%%%%%%%%%%%%%%%%%%%%%
 
Let us now focus on the optimization problem proposed above for the trade-off measure $\Gamma$. Hence we look for the maximum efficiency $\eta$ for a fixed output power $P$, i.e., we maximize $\Gamma$ for fixed values of $\delta_P$. Although this can be done analytically, the resulting formulas are quite involved and thus we present only numerical results. In Figs.~\ref{fig:universal2} and \ref{fig:universal3} the yellow dotted lines mark the maximum values of $\Gamma$ for fixed values of $\delta_{P}$ (which is the parameter of these curves) and for the individual values of the parameter $A$. Fig.~\ref{fig:universal4}(a) shows relative gain in efficiency $\delta_{\eta}$ corresponding to these maximum values. For small relative losses of the power incomparably larger relative gains of the efficiency are achieved. Consider for example the relative loss in power $\delta_{P} = -0.05$. Then the individual relative gains in efficiency are $\delta_{\eta} \approx 0.11$ for $A = \sqrt{0.1}$ (red solid line), $\delta_{\eta} \approx 0.16$ for $A=1$ (black dashed line) and $\delta_{\eta} \approx 0.2$ for $A = \sqrt{10}$ (blue dotted line), i.e., larger gains are obtained for larger $A$ values. 

Nevertheless, the performance of a heat engine is not determined by the variables $\delta_{P}$ and $\delta_{\eta}$, but rather by their absolute counterparts $P$ and $\eta$. In order to determine their values from Eqs.~(\ref{eq:reldP_def})-(\ref{eq:reldeta_def}) it is necessary to know the maximum power $P^{\star}$ (\ref{eq:P_opt}) and the corresponding efficiency $\eta^{\star}$ (\ref{eq:eta_opt}). While $\eta^{\star}$ depends solely on the parameters $A$ and $\eta_{\rm C}$ and thus the efficiency $\eta$ can be readily obtained, $P^{\star}$ is model-dependent and we calculate it again for the case of a particle diffusing in the log-harmonic potential (see Tab.~\ref{tab:potential_state}) considered in the preceding sections. Fig.~\ref{fig:universal4}(b)-(c) shows the resulting power and efficiency [the individual lines have the same meaning as in the panel (a)]. The overlapping lines for power plotted in Fig.~\ref{fig:universal4}(b) for  $10\mu_{\rm h}=\mu_{\rm c} =10$ ($A = \sqrt{0.1}$, red solid line) and for $\mu_{\rm h}=10\mu_{\rm c} =10$ ($A = \sqrt{10}$, blue dashed line) again demonstrate the symmetry of the maximum power (\ref{eq:P_opt}) with respect to the mobilities. While the power $P = P^{\star} + P^{\star} \delta_P$ grows with decreasing $\sqrt{A_{\rm I}} + \sqrt{A_{\rm III}}$ (see Eq.~(\ref{eq:P_opt})), the efficiency plotted in panel (c) favors engines characterized by small $A = \sqrt{A_{\rm III}/A_{\rm I}}$ values, i.e., in the case of diffusion-based heat engines, by small ratio of the mobilities, $\mu_{\rm h}/\mu_{\rm c}$. Altogether we can conclude that best performance can be expected from low-dissipation Carnot cycles characterized by small, but finite parameter $A_{\rm I}$ and by vanishing parameter $A_{\rm III}$ ($A_{\rm III} \to 0$).

%%%%%%%%%%%%%%%%%%%%%%%%%%%%%%%%%%%%%%%%%%%%%%%%%%%%%%%%%%%%%%%%%%
%%%%%%%%%%%%%%%%%%%%%%%%%%%%%%%%%%%%%%%%%%%%%%%%%%%%%%%%%%%%%%%%%%

In closing this section let us focus on the
optimization problem proposed for the trade-off measure $\Lambda$, i.e., maximization of $\Lambda$ as the function of $\tau$ and $a$. Also this can be done analytically, but the results are again rather complicated and it is simpler to find the solution numerically. We will restrict our considerations here to the totally asymmetric limits $A\to\infty$ and $A\to 0$ which lead to quite concise results. In the limit $A\to\infty$ we have $t_{\rm p}^{\star} \to \infty$ and $\alpha^{\star} \to 0$. On the other hand for $A \to 0$ we have $t_{\rm p}^{\star} \to \infty$  and $\alpha^{\star} \to 1$. The results obtained in these limits approximate well those for large and small values of $A$ where $t_{\rm p}^{\star} < \infty$, $0<\alpha^{\star} <1$ and $P^{\star} > 0$.

The limiting relative changes in power and efficiency for $A\to\infty$ read
\begin{eqnarray}
\lim_{A\to \infty}\delta_{P} &=& -\left(\frac{\tau}{1+\tau}\right)^2\,,
\label{eq:dP_Binf}\\
\lim_{A\to \infty}\delta_{\eta} &=& \frac{\tau}{1+\tau}\,.
\label{eq:deta_Binf}
\end{eqnarray}
Hence $\lim_{A\to \infty} \Lambda = (1 + 2 \tau)^2/(1 + \tau)^3$ and the maximum $\Lambda = 32/27$ is attained for an arbitrary $a$ at $\tau = 1/2$. The corresponding power $P = 8P^{\star}/9$ and efficiency $\eta = 4\eta^{\star}/3 = 2\eta_{\rm C}/3$ yield $\Gamma = 3$ and thus also in this limiting case the engine should not work at the maximum power (the relative loss in power $\delta_P = - 1/9$ is rather small as compared to quite large relative gain in efficiency $\delta_{\eta} = 1/3$).

For $A \to 0$ we have $a \in [-1,0]$ and it turns out that the maximal value of $\Lambda$ is achieved for $a\to 0$ (numerical result). For the relative changes in power and efficiency we obtain:
\begin{eqnarray}
\lim_{a,A\to 0}\delta_{P} &=&  -\left(\frac{\tau}{1+\tau}\right)^2\,,
\label{eq:dP_B0}\\
\lim_{a,A\to 0}\delta_{\eta} &=& 
\frac{2\tau}{2(1 + \tau) - \eta_{\rm C}}\frac{1}{1-\eta_{\rm C}}\,.
\label{eq:deta_B0}
\end{eqnarray}
Thus $\lim_{a,A\to 0} \Lambda = (1 + 2 \tau)^2 (2-\eta_{\rm C})/\{(1 + \tau)^2 [2 (1 + \tau)-\eta_{\rm C}]\}$ and the maximum $\Lambda$ is attained at $\tau = (\sqrt{9-8\eta_{\rm C}} - 1)/4$ and $a = 0$. Also in this case the corresponding trade-off $\Gamma = (3 + \sqrt{9-8\eta_{\rm C}})/2$, $\Gamma \in (2,3)$, favors other regimes than that of maximum power. The efficiency at maximum $\Lambda$ reads $\eta = (3-\sqrt{9-8\eta_{\rm C}})/2$.

Similarly as the EMP (\ref{eq:eta_opt}), the efficiency at maximum $\Lambda$ is a monotonically decreasing function of the parameter $A$ (numerical result, see also results shown for the diffusion-based heat engine in Fig.~\ref{fig:universal4}). Thus the
efficiency $\eta_- = 2\eta_{\rm C}/3$ obtained in the limit $A\to\infty$ represents the lower bound on the efficiency at maximum $\Lambda$ and, similarly, the efficiency $\eta_+ = (3-\sqrt{9-8\eta_{\rm C}})/2$ represents the respective upper bound. The efficiency at maximum trade-off $\Lambda \propto \eta P$ for low-dissipation heat engines thus must lie in the interval
\begin{equation}
\frac{2}{3}\eta_{\rm C} \le \eta \le \frac{3-\sqrt{9-8\eta_{\rm C}}}{2}\,.
\label{eq:BoundsMV}
\end{equation}
For comparison we review also other bounds obtained for efficiencies of low-dissipation heat engines corresponding to certain maximum parameters. The best known are bounds for the EMP
\cite{Esposito2010b}
\begin{equation}
\frac{1}{2}\eta_{\rm C} \le \eta \le \frac{\eta_{\rm C}}{2-\eta_{\rm C}}\,.
\label{eq:BoundsMP}
\end{equation}
Other interesting result from literature are bounds on efficiency at maximum value of the parameter $\dot{\Omega} = (2\eta - \eta_{\rm C})Q_{\rm h}/t_{\rm p}$ which provides a compromise between maximum work performed
and minimum work lost \cite{Tomas2013}. They read
\begin{equation}
\frac{3}{4}\eta_{\rm C} \le \eta \le
\frac{3-2 \eta_{\rm C}}{4-3 \eta_{\rm C}}\eta_{\rm C}\,.
\label{eq:BoundsMO}
\end{equation}

%%%%%%%%%%%%%%%%%%%%%%%%%%%%%%%%%%%%%%%%%%%%%%%%%
\begin{figure}
	\centering
		\includegraphics[width=1.0\columnwidth]{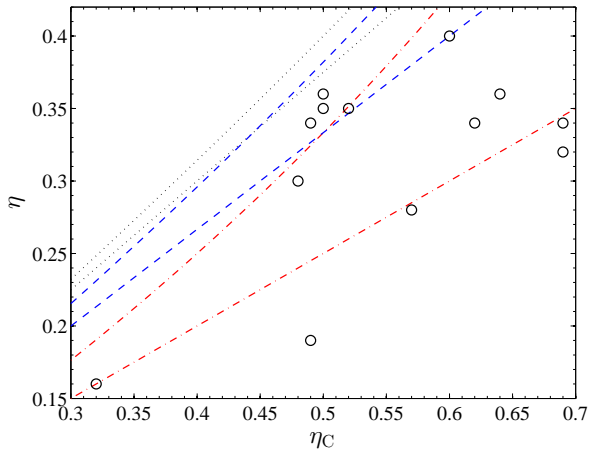}
\caption{(Color online) Comparison of the bounds (\ref{eq:BoundsMV})-(\ref{eq:BoundsMO}) on efficiency with experimental data. Blue dashed lines show the bounds for the efficiency at maximum trade-off $\Lambda$ (\ref{eq:BoundsMV}). Red dot-dashed lines and black dotted lines depict the same for the EMP (\ref{eq:BoundsMP}) and for the efficiency at maximum trade-off $\dot{\Omega}$ (\ref{eq:BoundsMO}) respectively. Black circles stand for measured efficiencies of various thermal plants, see Tab.~\ref{tab:data} \cite{Tomas2013, Esposito2010b}.}
	\label{fig:power_plants}
\end{figure}
%%%%%%%%%%%%%%%%%%%%%%%%%%%%%%%%%%%%%%%%%%%%%%%%%

%%%%%%%%%%%%%%%%%%%%%%%%%%%%%%%%%%%%%%%%%%%%%%%%%
\begin{table}[ht]%The best place to locate the table environment is directly after its first reference in text
\caption{\label{tab:data}%
Measured efficiencies $\eta$ of actual thermal plants taken from \cite{Tomas2013, Esposito2010b}. The Carnot efficiencies $\eta_{\rm C}$ are calculated from working temperatures of the individual power plants. The symbols $\eta_-$ and $\eta_+$ stand for the bounds (\ref{eq:BoundsMV}) on the efficiency at maximum trade-off $\Lambda\propto\eta P$.}
\begin{ruledtabular}
\begin{tabular}{l|l|l|l|l}
Thermal plant  & $\eta_{\rm C}$ & $\eta$ & $\eta_-$ & $\eta_+$ \\[1pt]
\colrule
\colrule
Doel 4, Nuclear, Belgium      & 0.50 & 0.35 & 0.33 & 0.38 \\[1pt]
Almaraz II, Nuclear, Spain    & 0.52 & 0.35 & 0.35 & 0.40 \\[1pt]
Sizewell B, Nuclear, UK       & 0.50 & 0.36 & 0.33 & 0.38 \\[1pt]
Cofrentes, Nuclear, Spain     & 0.49 & 0.34 & 0.33 & 0.37 \\[1pt]
Heysham, Nuclear, UK          & 0.60 & 0.40 & 0.40 & 0.48 \\[1pt]
West Thurrock, Coal, UK       & 0.64 & 0.36 & 0.43 & 0.52 \\[1pt]
CANDU, Nuclear, Canada        & 0.48 & 0.30 & 0.32 & 0.36 \\[1pt]
Larderello, Geothermal, Italy & 0.32 & 0.16 & 0.21 & 0.23 \\[1pt]
Calder Hall, Nuclear, UK      & 0.49 & 0.19 & 0.33 & 0.37 \\[1pt]
Steam/Mercury, USA            & 0.62 & 0.34 & 0.41 & 0.50 \\[1pt]
Steam, UK                     & 0.57 & 0.28 & 0.38 & 0.45 \\[1pt]
Gas Turbine, Switzerland      & 0.69 & 0.32 & 0.46 & 0.57 \\[1pt]
Gas Turbine, France           & 0.69 & 0.34 & 0.46 & 0.57 \\
\end{tabular}
\end{ruledtabular}
\end{table}
%%%%%%%%%%%%%%%%%%%%%%%%%%%%%%%%%%%%%%%%%%%%%%%%%

Although the low-dissipation assumption may not be fulfilled in actual thermal plants, it is tempting to confront the efficiency intervals (\ref{eq:BoundsMV})-(\ref{eq:BoundsMO}) with real data.
This is done in Fig.~\ref{fig:power_plants}, where we compare the individual efficiency intervals with experimental data measured for various actual thermal plants (Tab.~\ref{tab:data}, data taken from \cite{Tomas2013, Esposito2010b}).
Within the bounds for the efficiency at maximum $\Lambda$ fall five experimental points while within those at maximum power lie seven of them. The points belonging to the $\Lambda$ sector lie closer to its middle, while the points which fall into the maximum power sector lie closer to its borders. 
Hence this (very limited) set of data suggests that at least some real thermal plants operate close to optimum $\Lambda$ conditions rather than close to maximum power. Notably, the data which fits well into the interval corresponding to the optimal $\Lambda$ were all obtained from nuclear power plants, see Tab.~\ref{tab:data}.

%%%%%%%%%%%%%%%%%%%%%%%%%%%%%%%%%%%%%%%%%%%%%%%%%%%%%%%%%%%%%%%%%%%%%%%%%%%%%%%%%%%%%%%%%%%%%%%%%%%%%%%%%%%%%%%%%
\section{Conclusions and outlooks}
\label{sec:conclusion}

We have utilized the known results \cite{Esposito2010b, Schmiedl2008} on the EMP for low-dissipation Carnot cycles to
derive new universality in optimization of trade-off between power and efficiency. Our main results are Eqs.~(\ref{eq:reldP}) and (\ref{eq:reldeta}) which demonstrate that any trade-off measure expressible in terms of efficiency $\eta$ and the ratio $P/P^{\star}$ of power to its maximum value exhibits the same degree of universality as the EMP. Consequently, for many systems the trade-off measure can be optimized independently of most details of the dynamics and of the coupling to thermal reservoirs. 

In order to demonstrate this result we have introduced two specific trade-off measures. The first one (\ref{eq:tradeoff}) is designed for finding optimal efficiency for a given output power. The performed optimization clearly reveals diseconomy of engines working at maximum power. Engines operating in regimes near the maximum power with slightly smaller power and considerable larger efficiency are more economical (see Figs.~\ref{fig:universal2}-\ref{fig:universal4}). Within the second example we have derived universal lower and upper bounds (\ref{eq:BoundsMV}) on the efficiency at maximum trade-off given by the product of power and efficiency (\ref{eq:Lambda}). Comparison of the obtained bounds (\ref{eq:BoundsMV}) with data from actual nuclear power plants gives promising results, see Fig~\ref{fig:power_plants}. The data points outside the interval (\ref{eq:BoundsMV}) could result from thermal plants not operating at maximum value of the product of power and efficiency (\ref{eq:Lambda}) or not operating in the low-dissipation regime at all.

All the results have been illustrated on a diffusion-based heat engine specified in Secs~\ref{sec:model}-\ref{sec:max_eta}. These engines operate in the low-dissipation regime given that the used driving minimizes the work dissipated during the isothermal branches. The peculiarities of the corresponding optimization procedure have been reviewed and thoroughly discussed in Secs.\ref{sec:max_eta} and \ref{sec:maxP}. Let us note that this analysis has been performed for overdamped models only and it could be interesting to consider also the underdamped case. In this respect only few studies are available in the literature so far \cite{Rana2014, Rana2015,Benjamin2008,Tu2014}.

The low-dissipation assumption allows performing quite general analyzes. While this assumption may be reasonable for many systems \cite{Esposito2010b}, it is definitely not general. Therefore it is important to investigate also cases where the work dissipated during the isothermal branches behaves differently than in the low-dissipation regime. The present study has been performed for Carnot cycles with infinitely fast adiabatic branches. Similar research of Carnot cycles with adiabatic branches performed in finite-time \cite{Martinez2014} and also of other thermodynamic cycles like the Joule cycle or the Stirling cycle could provide further insight.

%%%%%%%%%%%%%%%%%%%%%%%%%%%%%%%%%%%%%%%%%%%%%%%%%%%%%%%%%%%%%%%%%%%%%%%%%%%%%%%%%%%%%%%%%%%%%%%%%%%%%%%%%%%%%%%%%%
%%%%%%%%%%%%%%%%%%%%%%%%%%%%%%%%%%%%%%%%%%%%%%%%%%%%%%%%%%%%%%%%%%%%%%%%%%%%%%%%%%%%%%%%%%%%%%%%%%%%%%%%%%%%%%%%%%
%merlin.mbs apsrev4-1.bst 2010-07-25 4.21a (PWD, AO, DPC) hacked
%Control: key (0)
%Control: author (72) initials jnrlst
%Control: editor formatted (1) identically to author
%Control: production of article title (-1) disabled
%Control: page (0) single
%Control: year (1) truncated
%Control: production of eprint (0) enabled
%

%%%%%%%%%%%%%%%%%%%%%%%%%%%%%%%%%%%%%%%%%%%%%%%%%%%%%%%%%%%%

\end{document}